\documentclass[aps,prl,reprint,longbibliography,floatfix]{revtex4-2}
\usepackage{graphicx}
\usepackage{bm}
\usepackage{amsmath}
\usepackage{amssymb}
\usepackage{comment}
\usepackage{soul}
\usepackage[utf8]{inputenc}
\usepackage[T1]{fontenc}
\usepackage[dvipsnames]{xcolor}
\usepackage{newunicodechar}
\usepackage{kotex}

\begin{document}
\title{Motility-Induced Pinning in Flocking System with Discrete Symmetry}

\author{Chul-Ung Woo}
\affiliation{Department of Physics, University of Seoul, Seoul 02504, Korea}
\author{Jae Dong Noh}
\affiliation{Department of Physics, University of Seoul, Seoul 02504, Korea}

\date{\today}

\begin{abstract}
{We report a motility-induced pinning transition in the active Ising model
for a self-propelled particle system with discrete symmetry. This model was
known to exhibit a liquid-gas type flocking phase transition, but a recent
study reveals that the polar order is metastable due to droplet excitation.
Using extensive Monte Carlo simulations, we demonstrate that, for an
intermediate alignment interaction strength, the steady state is
characterized by traveling local domains, which renders the polar order
short-ranged in both space and time. We further demonstrate that interfaces
between colliding domains become pinned as the alignment interaction
strength increases. A resonating back-and-forth motion of individual
self-propelled particles across interfaces is identified as a mechanism for
the pinning. We present a numerical phase diagram for the motility-induced
pinning transition, and an approximate analytic theory for the growth and
shrink dynamics of pinned interfaces. Our results show that pinned interfaces grow to a macroscopic size preventing the polar order in the regime where the particle diffusion rate is sufficiently smaller than the self-propulsion rate. The growth behavior in the opposite regime and its implications on the polar order remain unresolved and require further investigation.}
\end{abstract}

\maketitle
{\noindent\it Introduction} --
Active matter, consisting of self-propelled particles~(SPPs), displays intriguing collective phenomena~\cite{marchetti2013hydrodynamics,bechinger2016active}. SPPs, such as migrating cellular organisms~\cite{di2010bacterial,huber2018emergence,henkes2020dense}, swarming animals~\cite{buhl2006disorder,giardina2008collective,ballerini2008interaction,ward2008quorum,cavagna2014bird,cavagna2017dynamic}, synthetic materials~\cite{bricard2013emergence,kumar2014flocking,yan2016reconfiguring,liebchen2018synthetic}, and so on, convert internal or external energy into kinetic motion. The self-propulsion distinguishes active matter from thermal equilibrium systems. When SPPs interact through local velocity alignment, they can exhibit long-range polar order with broken continuous symmetry, even in two dimensions~\cite{vicsek1995novel,toner1995long}. This is in contrast to thermal equilibrium systems, where the Mermin-Wagner theorem prohibits such ordering~\cite{mermin1966absence}. Repulsive interactions among SPPs can lead to a phase separation, whereas attractive interactions would be necessary in thermal equilibrium systems~\cite{fily2012athermal,buttinoni2013dynamical,cates2015motility,Caprini.2020te}. Active matter systems incorporating other elements, such as discrete symmetry~\cite{csahok1995lattice,solon2013revisiting,solon2015flocking,mangeat2020flocking,chatterjee2020flocking,dittrich2021critical,solon2022susceptibility,chatterjee2022polar}, multiple species~\cite{Menzel.2012,fruchart2021non,chatterjee2023flocking}, and quenched disorder~\cite{Chepizhko.2013,Ro.2021,duan2021breakdown,Vahabli.2023}, have been attracting growing interest.

The Vicsek model is a well-established model for the flocking transition~\cite{vicsek1995novel}. It comprises SPPs whose self-propulsion direction is represented by a continuous XY spin variable. Both a field-theoretic renormalization group study~\cite{toner1995long,toner1998flocks} and extensive numerical simulations confirm that the model indeed exhibits the long-range polar order~\cite{gregoire2004onset,bertin2006boltzmann,chate2008collective,chate2008modeling,bertin2009hydrodynamic,baglietto2009nature,ihle2013invasion}. 
{The active Ising model~(AIM) is a discrete version of the Vicsek model~\cite{solon2013revisiting,solon2015flocking}.}
As a discrete model, it enables large scale numerical simulations and facilitates an analytically tractable { hydrodynamic theory~\cite{bertin2006boltzmann,Kourbane-Houssene.2018,Scandolo.2023}}. The system has been known to exhibit a liquid-gas type phase transition between a disordered~(gas) phase and a polar ordered~(liquid) phase. These phases are separated by a coexistence phase where particles are phase separated into gas and liquid regions. The liquid region forms a macroscopic band traveling over the gaseous background. The liquid-gas transition picture has also been confirmed in the Vicsek model with a modification that macrophase separation is replaced with microphase separation in the coexistence phase~\cite{solon2015phase}.

Recent studies have revealed the fragility of the long-range polar order in active matter systems. In the Vicsek model, a single point-like obstacle or a finite counter-propagating blob can disrupt the global polar order~\cite{codina2022small}. The AIM exhibits even greater fragility~\cite{Benvegnen.2023ta}. When the system is prepared in an ordered state, droplets spontaneously nucleate and counter-propagate at a constant speed, ultimately destroying the initial polar order. 
These findings raise an interesting question: does the long-range polar order truly exist in the active matter systems with discrete symmetry? If not, what would be the asymptotic phase?

This Letter addresses those questions specifically in the context of the
AIM. We demonstrate numerically that both the ordered and the coexisting
states in the liquid-gas transition scenario ultimately evolve to a state
consisting of randomly oriented finite-size traveling droplets, rendering
the polar order short-ranged in both space and time. Interestingly, we
discover as the alignment interaction strength increases further, the system
undergoes a {\em motility-induced pinning}~(MIP) transition instead of the
expected liquid-gas transition. In the pinned phase, interfaces between
domains oriented in opposite directions become immobile. Individual
particles accumulate near these interfaces and exhibit a back and forth
resonating motion.  We will provide a numerical evidence for the MIP and its
analytic justification, and discuss the implication of the MIP on the polar
order in the last section.
\begin{figure}
    \includegraphics[width=\columnwidth]{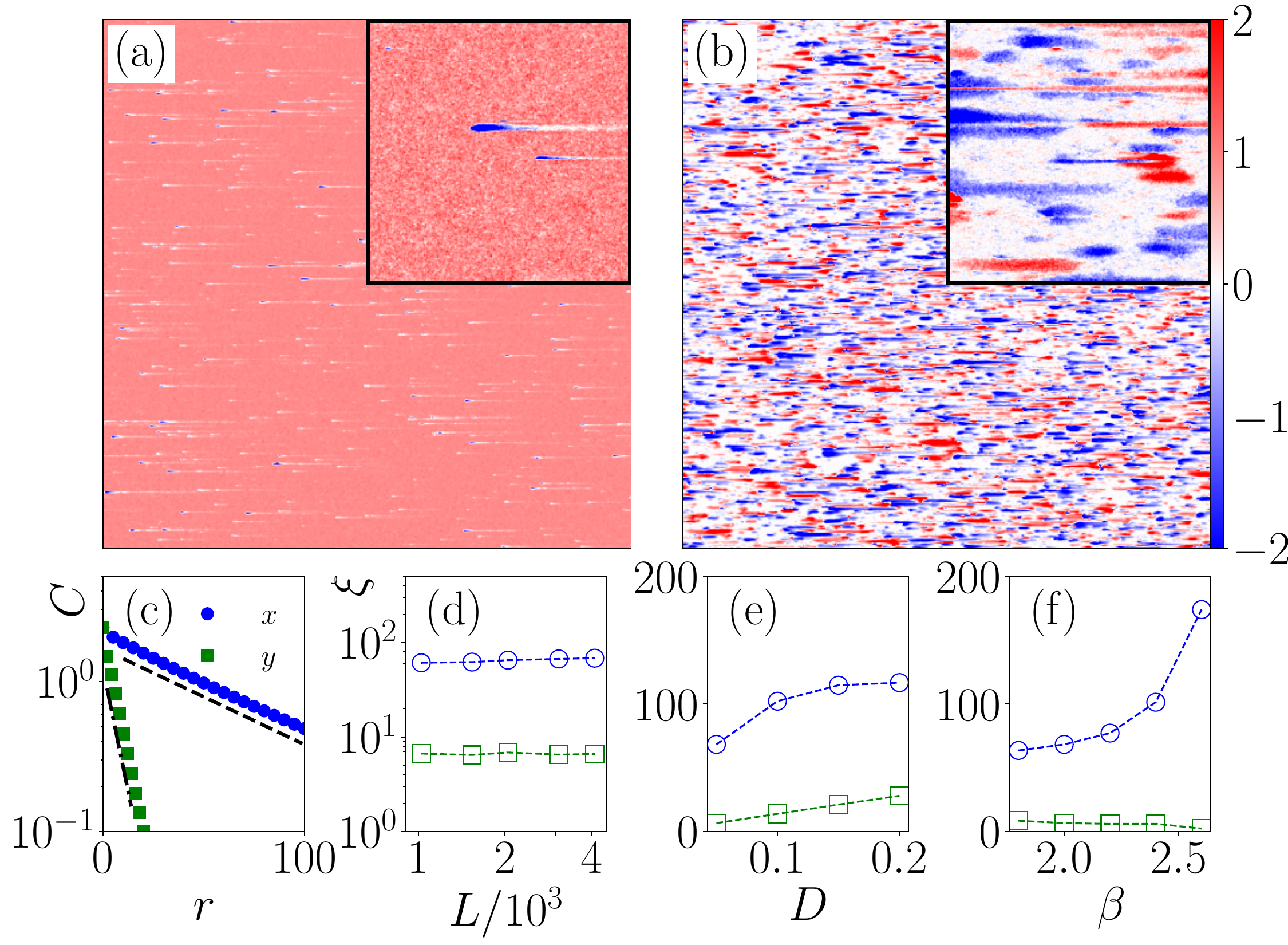}
    \caption{Snapshots taken at $4\times 10^3$ MCS in (a) and $5\times 10^4$ MCS in (b) of an initially ordered system of size $4096^2$. The insets present an enlarged view of a subsystem of size $256^2$. The normalized magnetization $m/\rho_0$ is color-coded according to the color bar. (c) Magnetization correlation functions in the steady state along the $x$ and $y$ directions (symbols) following an exponential decay~(dashed lines). [Parameters: $L = L_x = L_y = 4096$, $\rho_0 = 8$, $D=0.05$, $v=1$, and $\beta=2$] Also shown is the dependence of $\xi_x$~(circles) and $\xi_y$~(squares) on $L$ in (d), $D$ in (e), and $\beta$ in (f).}
    \label{fig1}
\end{figure}

{\noindent\it Active Ising model} --
The AIM~\cite{solon2013revisiting} comprises $N$ SPPs on a two-dimensional~(2D) lattice of $L_x\times L_y$ sites with periodic boundary conditions. The overall density is denoted as $\rho_0 = N/(L_x L_y)$. Each particle is associated with an Ising spin variable, $s=\pm 1$, indicating its self-propulsion direction. Particles hop to one of their four neighbors at diffusion rate $4D$, self-propel to a neighboring site on the right~($s=+1$) or left~($s=-1$) at rate $v$, and can flip their spin state~($s\to -s$) at rate $w e^{-\beta s m_{\bm r}/\rho_{\bm r}}$, where $\rho_{\bm r}$ and $m_{\bm r}$ denote the number of particles and the magnetization at the residing site $\bm{r}=(x,y)$, respectively. The ratio $p_{\bm r} = m_{\bm r}/\rho_{\bm r}$ is called the polarization. We will set $w=1$. The parameter $\beta$, called the inverse temperature, represents a strength of an alignment interaction of self-propulsion directions. The polar order manifests itself as a ferromagnetic order.

The AIM can be described by the continuum hydrodynamic equation, based on a local mean-field approximation, for the density field $\rho = \rho(\bm{r}, t)$ and the magnetization field $m = m(\bm{r}, t)$~\cite{solon2015flocking,Kourbane-Houssene.2018,Scandolo.2023,Benvegnen.2023ta}:
\begin{equation}
\begin{split}\label{eq:hydrodynamic}
\partial_t \rho &= \nabla \cdot \mathsf{D} \nabla \rho -v \partial_x m \\
\partial_t m &= \nabla \cdot \mathsf{D} \nabla m -v \partial_x \rho + F(\rho,m)
\end{split}
\end{equation}
Here, $F(\rho,m) = 2\rho \sinh{(\beta m/\rho)} - 2m \cosh{(\beta m/\rho)}$, and the diffusion matrix $\mathsf{D}$ is diagonal with elements $D_{x} = D+v/2$ and $D_{y} = D$. 

We have simulated the AIM dynamics using a parallel update Monte Carlo~(MC) method. During one MC sweep~(MCS), corresponding to a time interval $\Delta t = 1/(4D+v+e^{\beta})$, all particles attempt hopping, self-propulsion, or spin flip in parallel~\footnote{We confirmed that the results are qualitatively the same under the parallel update and the random sequential update.}.

{\it\noindent Metastability of ordered and coexistence phases} --
We confirm the metastability of the ordered state~(see Fig.~\ref{fig1}), which was first reported in Ref.~\cite{Benvegnen.2023ta}. The system, starting from an ordered initial state, evolves into a state with multiple traveling droplets, nucleated spontaneously. These droplets grow and merge into larger ones. At the same time, they also suffer from spontaneous nucleation of droplets of opposite polarization, and break up into smaller pieces. This competition between growth and break-up drives the system to a steady state with randomly distributed local domains~\footnote{See Supplemental Material.}. 

The characteristic size of droplets is estimated using the correlation function $C(\bm{r}) := \sum_{\bm{r}_0} \langle m_{\bm{r}+\bm{r}_0} m_{\bm{r}_0} \rangle_{ss} / (\rho_0^2 L_x L_y)$, where $\langle \rangle_{ss}$ denotes a steady state time average. As shown in Fig.~\ref{fig1}(c), it decays exponentially with distance in both $x$ and $y$ directions. Importantly, the characteristic sizes $\xi_x$ and $\xi_y$ converge to finite values as the system size increases (Fig.~\ref{fig1}(d)). The characteristic sizes vary smoothly with other parameters like $D$ and $\beta$ (Fig.~\ref{fig1}(e, f)). These observations collectively indicate that the polar order in the steady state is short-ranged in both space and time.

The coexistence phase is also found to be metastable~\cite{Note2}. A droplet nucleates spontaneously inside a band and grows to a size $\xi = O(L_x)$ before escaping into the disordered background. Continuing its ballistic movement, it evaporates as constituent particles diffuse away. Notably, this diffusive evaporation would take a time $\tau_{\rm evap} \propto \xi^2$, while collision with the band takes $\tau_{\rm col} \propto L_x$. Therefore, in sufficiently large systems, the droplet repeatedly invades the band, ultimately destroying it. {In the infinite system, a liquid band itself is macroscopic, hence metastable as the liquid state is.}

{\it Motility-induced pinning} -- 
Numerical results suggest that the AIM exhibits only a crossover {to a locally ordered state}, not a sharp liquid-gas transition. The global polar order is hindered by the droplet excitation~\cite{Benvegnen.2023ta}. We will demonstrate, however, that these traveling droplets cease to exist as the inverse temperature $\beta$ increases further, and the system undergoes a MIP transition. As $\beta$ increases, two distinct time scales emerge: one for spin flips~(fast process with rate $\sim e^\beta$) and the other for particle motion~(slow process with rate $4D+v$). This separation of time scales significantly impacts droplet dynamics. 

\begin{figure}
    \includegraphics[width=\columnwidth]{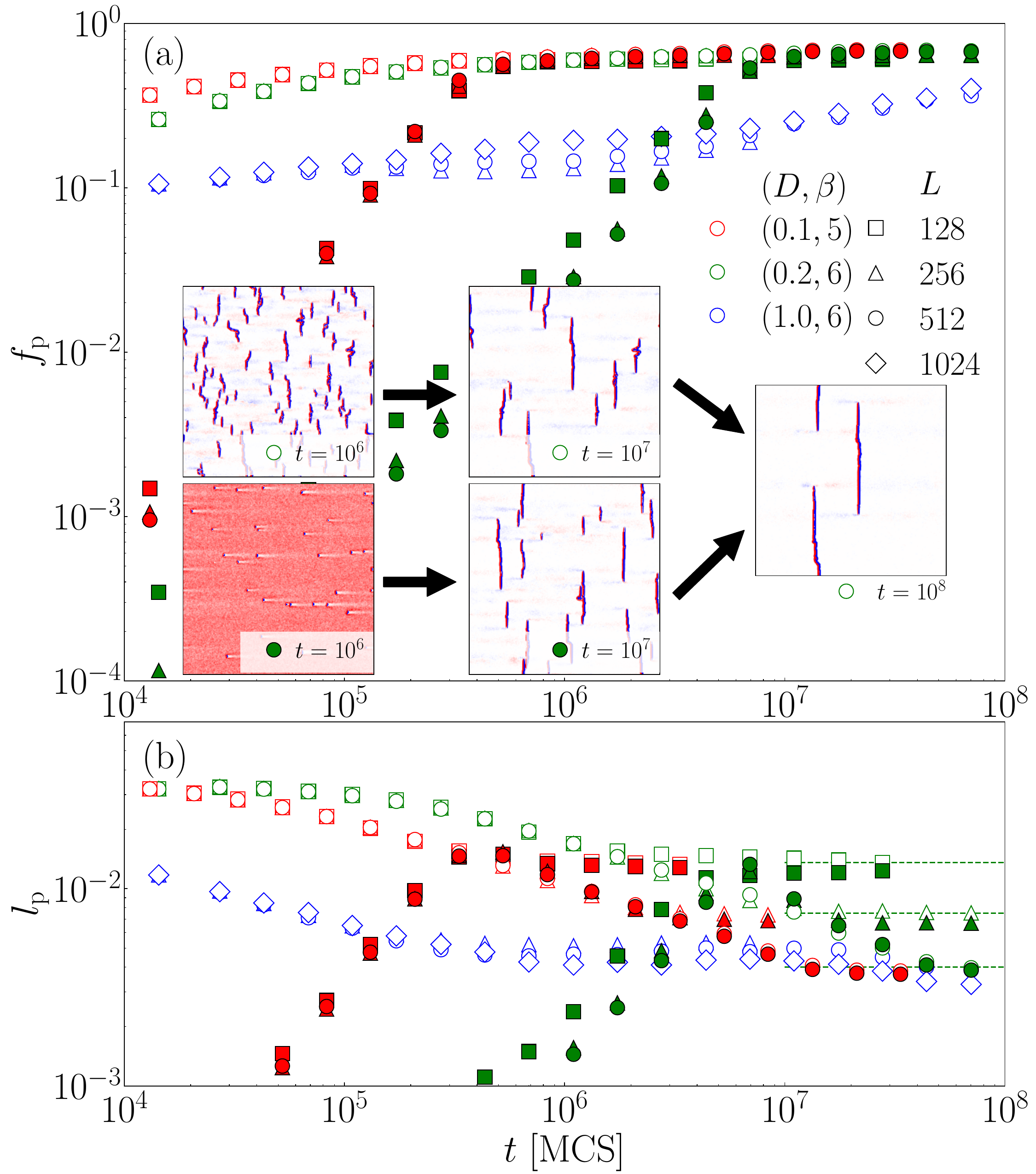}
    \caption{$f_{\rm p}$ and $l_{\rm p}$ under a RIC~(empty symbols)
            and an OIC~(filled symbols) for a few values of $(D, \beta)$ and $L_x=L_y=L$ as
            marked in Fig.~\ref{fig3}(a)~[Parameters: $\rho_0=4$ and $v=1.0$]. Panels in (a) illustrate typical magnetization configurations, color-coded as in Fig.~\ref{fig1}, in the transient, coarsening, and steady state regimes at $t=10^6$, $10^7$, and $10^8$MCS, respectively, from both initial conditions when $L=512$ and $(D,\beta)=(0.2, 6.0)$~\cite{Note2}. The dashed lines in (b) indicate the scaling $l_{\rm p}=O(L^{-1})$ at $(D,\beta)=(0.2, 6.0)$. 
    }\label{fig2}
\end{figure}

Consider the limiting case $\beta\to\infty$. When a particle moves to a nonempty site, the particle's spin instantly aligns with the polarization at the target site.
Consider two local domains of opposite polarization confronting each other with a domain wall or an interface separating them. Particles near the interface exhibit a back-and-forth oscillation: whenever a particle self-propels across the interface, it flips its self-propulsion direction and returns, and this process repeats. Consequently, the interface will be pinned in space. In analogy to the resonance structures found in chemical bonds~\cite{Muller.1994}, this oscillating motion will be called resonance. When $v=0$ (without self-propulsion), the resonance disappears, and the interfaces are never pinned. 
In 1D, a pinned interface becomes a point defect~\cite{Benvegnen.2022}. 

{The MIP phenomenon persists for finite values of $\beta$, as demonstrated in Fig.~\ref{fig2}. The pinning is characterized by $f_{\rm p}$, the fraction of particles trapped in pinned interfaces~(PIs), and $l_{\rm p}$, the total length of PIs per site. Under a random initial condition~(RIC), PIs are created as randomly distributed domains collide with each other. The system undergoes a coarsening until reaching 
the state with $f_{\rm p}=O(1)$ and $l_{\rm p} = O(L_x^{-1})$~\footnote{The finite size scaling behavior $l_{\rm p}=O(L_x^{-1})$ has been confirmed in rectangular systems with $L_x\neq L_y$.}, wherein macroscopic PIs span the lattice in the $y$ direction.
{This state will be referred to as a macroscopic PI state~(MPIS).} Under an ordered initial condition~(OIC), the 
{MPIS} is accessible numerically for small values of $D$. In a short time regime, $f_{\rm p}$ increases super-linearly indicating that PIs are nucleated at a constant rate and their size grows. Then, the initial condition dependence disappears at a system-size-independent characteristic time and the system enters the coarsening regime to reach the 
{MPIS}. These results suggest that the 
{MPIS} constitutes a stable steady state phase.

A few remarks are in order. (i)~Snapshots in Fig.~\ref{fig2} reveal a particle flux among PI segments. Individual resonating particles diffuse along a PI segment. Reaching an edge, they leak out and flow until being trapped in another segment. Due to this leak current, short PI segments blocked by longer ones 
shrink while longer ones grow during the coarsening~\cite{Note2}. 
(ii)~The system under the OIC could reach the 
{MPIS} within the simulation time scale only for small $D$. The relaxation dynamics is initiated by spontaneously nucleated and pinned droplets. Their nucleation time is known to increase exponentially with $D$~\cite{Benvegnen.2023ta}. We attribute the numerical difficulty to the rapid increase of the nucleation time. 
The coarsening dynamics becomes also slow as $D$ increases. At $D=1$, the systems with $L\gtrsim 1024$ have not reached the 
{MPIS} even at $10^8$ MCS~\cite{Note2}.

\begin{figure}
    \includegraphics[width=\columnwidth]{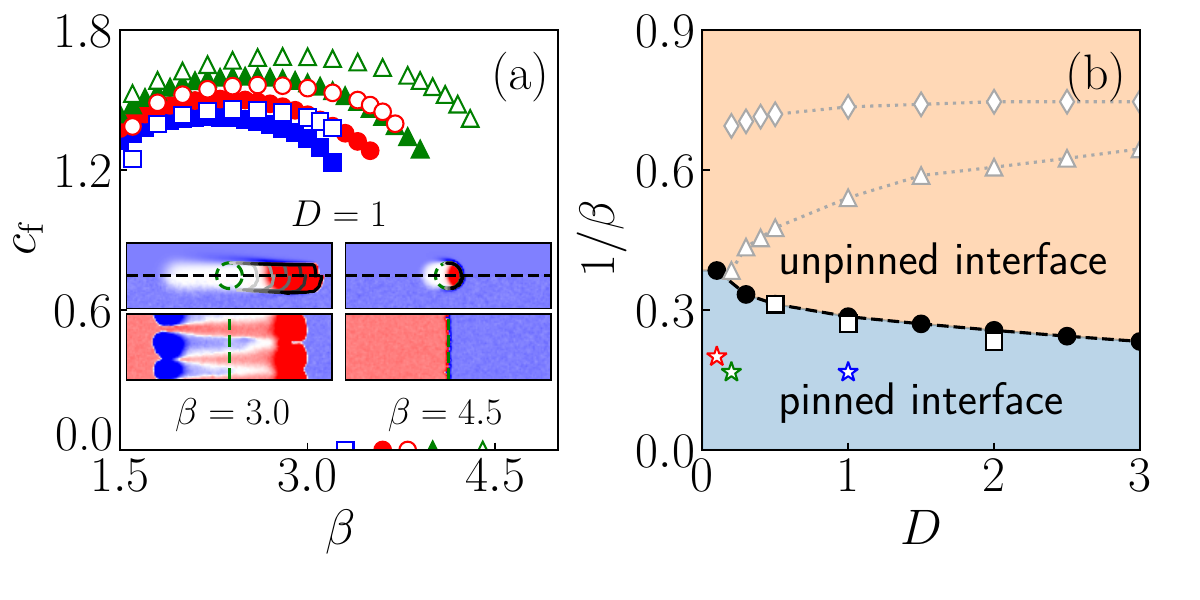}
    \caption{(a) Propagation speed of a droplet from the MC
    simulations~(empty symbols) and the hydrodynamic equation~(filled
symbols) [Parameters: $\rho_0=4$, $v=1$, $D=0.5$~(square), $1.0$~(circle),
and $2.0$~(triangle)]. The insets depict a droplet at successive time steps
from MC simulations (bottom half) and the hydrodynamic equation (top half).
Snapshots of a spanning interface are also compared. (b) Phase diagram at
$\rho_0=4$ and $v=1$ with transition points marked with empty symbols~(MC
simulations) and filled symbols~(hydrodynamic equation). The MIP transition
line is compared with the metastable liquid-gas transition lines~(dotted
lines){, which are estimated numerically from the binodal densities of metastable bands}. } \label{fig3}
\end{figure}

Interface dynamics is useful in investigating the MIP transition. Consider a spanning interface at $x=L_x/2$ separating the system into two halves. The interface roughens as domains penetrate into the other for $\beta < \beta_c$ while it is frozen for $\beta > \beta_c$~\cite{Note2}. 
To pinpoint the transition point $\beta_c$, we introduce an isolated circular droplet into an ordered state and measure the speed $c_{\text f}$ of the front~\cite{Note2}. As shown in Fig.~\ref{fig3}(a), $c_{\rm f}$ from MC simulations and from numerical solutions of the space-discretized hydrodynamic equations~\eqref{eq:hydrodynamic} exhibits a discontinuous jump to zero at $\beta=\beta_c$. The resulting phase diagram is shown in Fig.~\ref{fig3}(b). The speed and the phase diagram from both methods are in a qualitative agreement.
}
 
{\it MIP from hydrodynamic theory} --
The hydrodynamic theory of Ref.~\cite{Benvegnen.2023ta} predicts a traveling droplet solution only. We revisit the hydrodynamic theory to reconcile it with the phenomenon of MIP. Consider a droplet of density $\rho_{\text d}$ and magnetization $m_{\text d} = p \rho_{\text d}>0$, whose front moves at a constant speed $c_{\text f}$ on an ordered background characterized by particle density $\rho_{\text o}$ and magnetization $m_{\text o} = -p \rho_{\text o} < 0$. The polarization $p>0$ is determined by $p = \tanh \beta p$. Mass conservation requires that 
$c_{\text f} = \frac{m_{\text d} + |m_{\text o}|}{m_{\text d} - |m_{\text o}|} p v = \frac{\rho_d + \rho_{\text o}}{\rho_d - \rho_{\text o}} p v$. 
Combined with a small density gradient expansion, the hydrodynamic equation leads to a simplified description for the magnetization $m(z)$ along the symmetry axis of a droplet~\cite{Caussin.2014,Solon.2015604,Benvegnen.2023ta}:
\begin{equation}
    {D}_{x} \frac{d^2 m}{dz^2} = -\gamma(c_{\text f}) \frac{dm}{dz} - \frac{dV(m;c_{\text f})}{dm},
    \label{eq:NewtonMap}
\end{equation}
where $z = x - c_{\text f} t$ is the coordinate in the comoving frame, $\gamma(c_{\text f}) = c_{\text f} - v^2/c_{\text f}$, and $V(m;c_{\text f}) = \int F(\rho_{\text o}+v(m-m_{\text o})/c_{\text f}, m) dm$~[see Ref.~\cite{Benvegnen.2023ta} for  derivation]. Regarding $m$ as a coordinate and $z$ a time, this equation describes a damped motion of a particle of mass $D_{x}$ under the inverted double-well potential $V(m; c_{\text f})$ having two unstable fixed points at $m=m_{\text o}$ and $m_{\text d}$.
The traveling droplet solution corresponds to the heteroclinic orbit connecting these two unstable fixed points. The condition that the excess potential energy $\Delta V := V(m_{\text d};c_{\text f}) - V(m_{\text o}; c_{\text f})$ is exactly dissipated by the damping term determines $c_{\rm f}$ self-consistently.

When $\beta$ and $c_{\text f}$ are large, the self-consistent equation admits an approximate solution. Due to mass conservation, we have $\Delta \rho := \rho_{\text d}-\rho_{\text o} = 2vp\rho_{\text o}/c_{\text f} + O(c_{\text f}^{-2})$ and $\Delta |m| := m_{\text d} - |m_{\text o}| = 2vp^2\rho_{\text o}/c_{\text f} +  O(c_{\text f}^{-2})$. We can set the polarization $p=1$ neglecting an $O(e^{-2\beta})$ correction. Keeping only the leading order term, we obtain  
\begin{equation}
    \Delta V = \int_{m_{\text o}}^{m_{\text d}} F dm = \frac{6 v\rho_{\text o}^2}{\beta^2 c_{\text f}} e^\beta \left(1 + O(\beta^{-1}, c_{\text f}^{-1})\right).
\end{equation}
The energy dissipation $E_d = \frac{1}{2}\int_{-\infty}^{\infty} \gamma |m'(z)|^2 dz$ is approximated as $E_d \simeq \gamma (m_{\text d} - m_{\text o})^2 / \Delta z$, where $\Delta z$ denotes a time required for a transition from $m_{\text d}$ to $m_{\text o}$. 
For $m=m_{\rm o}+\delta m$ with $|\delta m| \ll 1$,
Eq.~\eqref{eq:NewtonMap} becomes
\begin{equation}
    D_x \delta m'' \simeq - \gamma \delta m' + k \delta m
\end{equation}
with a stiffness constant
$k = -\left.\frac{dF}{dm}\right|_{m=m_{\text o}} \simeq e^{\beta}$,
which yields that $\delta m \sim e^{-z/\tau_z}$ with
\begin{equation}
\tau_z = \frac{2D_x}{\gamma + \sqrt{\gamma^2 + 4 D_x k} }.
\end{equation}
Therefore, the transition time is given by $\Delta z = a \tau_z$ with a constant $a=O(1)$. Equating $\Delta V$ and $E_d$, we finally obtain 
\begin{equation}
    c_{\text f} \simeq \left(3av\sqrt{D_x}/2\right)^{1/2} \beta^{-1} e^{\beta/4} .
    \label{eq:cf_analytic}
\end{equation}
The exponential dependence in $\beta$ is consistent with the numerical solution of Eq.~\eqref{eq:NewtonMap}~(not shown here).

The resulting density gradient $\Delta\rho/\Delta z \sim e^{\beta/4}$ grows exponentially with $\beta$, which makes the density gradient expansion worse and worse as $\beta$ increases. We will argue that the traveling droplet solution breaks down beyond a threshold $\beta_c$ using the resonance mechanism. We consider a simplified model consisting of only two sites $A$ and $B$, representing locations on the left and right of the droplet's front interface, respectively.
During a time interval $\tau_{f} = e^{-\beta}$,  $f_+ = \rho_{\text d} (D+v)\tau_f$ particles of positive spin flow from $A$ to $B$, while $f_- = \rho_{\text o}(D+v)\tau_f$ particles of negative spin from $B$ to $A$. We choose a specific value $\tau_f = e^{-\beta}$ so that spin flips are suppressed during this time interval. The droplet can move forward only if the invading particles~($f_+$) outnumber the remaining particles~($\rho_{\text o}-f_-$) at $B$. Otherwise, the invaders reverse their spin state and return back to site $A$, starting the resonating motion. Therefore, the traveling droplet solution requires $f_+ > \rho_{\text o} - f_-$, which imposes an upper bound for $c_f$:
\begin{equation}
    c_f < v / \left(1-2(D+v) e^{-\beta}\right).
    \label{eq:cf_bound}
\end{equation}
For large $\beta$, this bound decreases whereas the solution in Eq.~\eqref{eq:cf_analytic} increases with $\beta$. This inconsistency implies the existence of a threshold value $\beta_c$, beyond which the resonance sets in and interfaces become pinned.  

{\it Coarsening dynamics of pinned interfaces} -- 
{To shed a light on the coarsening dynamics of PIs, we develop a theory governing the time evolution of the length of PI segments. It is analytically tractable when $D/v\ll 1$ and $\beta \gg 1$. We will keep only the leading order term neglecting $O(l^{-1}, D/v, e^{-\beta})$ corrections in this section.

Consider a PI segment of length $l$ in a homogeneous background of $+$ spins. Particles of total flux $F = \rho_0 vl$ land on the PI,
jump across it at rate $D+v$, diffuse along it at rate $D$, 
and leak away at both end. Assuming that $D/v \ll 1$, we model the PI segment as two unit-width columns of $+$ spins and $-$ spins with the leak current channels of unit width at both ends. 
Let $P$~($M$) be the terminal site of the positive~(negative) domain, in one end, 
and $P'$~($M'$) be the neighboring site through which leaking $+$ particles flow. Flux balance conditions yield that the mean occupation numbers at those sites are given by  $N_P \simeq N_M \simeq \rho_0 v l/(4D)$, $N_{P'} \simeq \rho_0 l/4$, and $N_{M'} \simeq \rho_0 l/2$.

The PI segment can {\em grow} by one lattice unit for a time interval $\tau_f$ if $-$ particles hopping from $M$ outnumber $+$ particles residing at $M'$. For large $l$, the number of hopping particles $n$ follows a Gaussian distribution of mean $\langle n\rangle = N_M D\tau_f$ and variance $\sigma_n^2 = N_M D\tau_f(1-D\tau_f)$. Thus, the growth probability is given by $P_g \simeq \frac{1}{\sqrt{2\pi}}\int_{z_g}^\infty e^{-z^2/2} dz\simeq \frac{1}{\sqrt{2\pi z_g^2}}e^{-z_g^2/2}$ with $z_g := (N_{M'}-\langle n\rangle)/\sqrt{\sigma_n^2} \simeq \sqrt{\frac{\rho_0 l}{v \tau_f}}$. On the other hand, it can {\em shrink} by one lattice unit if $+$ particles jumping from $P$ outnumber $-$ particles at $M$. Using the similar Gaussian statistics, we find that the shrink probability is given by $P_s \simeq \frac{1}{\sqrt{2\pi z_s^2}}e^{-z_s^2/2}$ with $z_s \simeq \sqrt{\frac{\rho_0 l}{4D \tau_f}}$. Thus, the rate $W = P/\tau_f$ follows the exponential scaling law 
\begin{equation}\label{exp_scaling}
    W(l) \sim e^{-l/l_0}
\end{equation}
for large $l$, where $l_0 = l_g \simeq 2 v \tau_f /\rho_0$ for growth and $l_0 = l_s \simeq 8D \tau_f /\rho_0$ for shrink. 
The growth is dominant~($l_g > l_s$) for small $D/v$ region, which explains
the emergence of the MPIS observed in Fig.~\ref{fig2}. The exponential
scaling law and the dominance of the growth~($l_g > l_s$) have been
confirmed numerically for $D/v \leq 1.0$ at $\beta=5$~\cite{Note2}. }

{\it Discussions} --
We have discovered a novel phenomenon of MIP in the AIM. Pinning typically occurs due to defects or impurities~\cite{Chepizhko.2013,Peruani.2018,Forgacs.2021,Sar.2023,Vahabli.2023}.
Our work reveals a unique mechanism for pinning in the absence of quenched disorder: a resonating back-and-forth motion of SPPs.
We have also studied the active $p$-state clock model, in which the
self-propulsion direction is modeled with $p$-state clock spins. We observed
the MIP phenomenon for $p\leq 4$~\cite{Note2}, which will be presented elsewhere.

Contrary to early expectations~\cite{solon2013revisiting,solon2015flocking,solon2015phase}, 
the AIM does not exhibit the liquid-gas transition due to the droplet excitation~\cite{Benvegnen.2023ta}. Instead, it displays the MIP transition. The nature of the asymptotic steady state below the MIP transition is determined by the growth and shrink dynamics of PIs. Our numerical and analytic results have shown that the growth is dominant and the system relaxes to the MPIS for small $D/v$ region. On the other hand, the spontaneous nucleation of droplets~\cite{Benvegnen.2023ta} and the coarsening dynamics of PIs become extremely slow as $D$ increases, which hinders probing the nature of the steady state in the whole region below the pinning transition. Since the competition between growth and shrink can have a profound impact on the global polar order, it calls for further studies especially in the large $D/v$ region, which is left for future work.

\begin{acknowledgments}
We acknowledge fruitful discussions with Heiko Rieger, Sunghan Ro, and Yarif Kafri.
This work is supported by the National Research Foundation of Korea (NRF) grant funded by the Korea government (MSIT) (No. RS-2024-00348526).
We also acknowledge the computing resources of Urban Big data and AI Institute (UBAI) at the University of Seoul.
\end{acknowledgments}
\bibliography{paper_rev3}

\renewcommand{\theequation}{S\arabic{equation}}
\renewcommand{\thefigure}{S\arabic{figure}}
\setcounter{equation}{0}
\setcounter{figure}{0}
\setcounter{secnumdepth}{2}
\onecolumngrid
\newpage
\begin{center}
{\bf\large Supplemental Materials for \\
``Motility-Induced Pinning in Flocking System with Discrete Symmetry''}\\
\vspace{4mm}
\setcounter{page}{1}

Chul-Ung Woo and Jae Dong Noh\\
\vspace{2mm}
{\it Department of Physics, University of Seoul, Seoul 02504, Korea}\\
\end{center}

\appendix

\section{Supplementary animations}

\begin{itemize}
    \item{File \texttt{1\_droplets\_OIC.mp4} corresponds to the supplementary movie for Fig.~1. It shows a time evolution of the microscopic model from a homogeneous ordered state to a state with nucleated droplets. Parameters: $L_x = L_y = 4096, \rho_0 = 8, \beta = 2, v=1, D=0.05$.}
    
    \item{File \texttt{2\_droplets\_band.mp4} corresponds to the supplementary movie showing a time evolution of the microscopic model from a macrophase separated state to a state with nucleated droplets. Parameters: $L_x = 2048, L_y=256, \rho_0=3, \beta=2, v=1, D=0.1$.}

    \item{Files \texttt{3\_MIP\_RIC.mp4} and \texttt{4\_MIP\_OIC.mp4} correspond to the supplementary movies for the insets of Fig. 2(a). These videos demonstrate the time evolution from the random initial condition and ordered initial condition, respectively, into the motility-induced pinning phase in the microscopic model. Parameters: $L_x=L_y=512, v=1, D=0.2, \beta=6, \rho_0 = 4$.}

    \item{File \texttt{5\_MIP\_RIC\_largeL.mp4} corresponds to the supplementary movie for Fig. 2. It demonstrates a slow relaxation dynamics to the motility-induced pinning phase in the microscopic model. Parameters: $L_x=L_y=4096, v=1, D=1, \beta=5, \rho_0 = 4$.}

    \item{Files \texttt{6\_droplet.mp4} and \texttt{7\_droplet\_DHE.mp4} are the supplementary movies for the insets of Fig. 3 (a). They show the time evolution of an isolated circular droplet of radius $r_0 = 50$ in the ordered background in the microscopic model and the discretized hydrodynamic equation, respectively. Parameters: $L_x=1024, L_y=256, v=1, D=1, \rho_o=4$.}

    \item{File \texttt{8\_ACM\_droplet.mp4} is the supplementary movie for the last paragraph of the main text. It shows the motion of pinned and unpinned droplets in the active $p$-state clock model with $p=4$~(left column) and $p=8$~(right column) in the microscopic model~(top row) and in the discretized hydrodynamic equation~(bottom row). Parameters: $L_x=L_y=1024, \rho_o = 10, \beta=6, v=1, D=1$.}
\end{itemize}

\section{Growth and shrink rates of a pinned interface}

We examined numerically the exponential scaling law $W_{g,s}(l) \sim e^{-l/l_{g,s}}$
for the growth and shrink rates of a PI derived in the main text. When
$\beta > \beta_c$, the rates were measured in the following way:
\begin{enumerate}
    \item  The system is prepared to be in an ordered state with a PI of length $l$ being implanted. Namely, all lattice sites are occupied by $+$ particles of mean density $\rho_0$ except for those within a $(2\times l)$ rectangle. They are occupied by $\rho_0'~( > \rho_0)$ particles whose spin states are $+$ in the $(1\times l)$ column on the left and $-$ in the $(1\times l)$ column on the right.

    \item A PI is identified as an interface between positively and
        negatively polarized domains. The occupation numbers at sites along
        the domain boundary are required to be larger than a cutoff value $\rho_{\rm cutoff} = 5\rho_0$.

    \item During simulations up to $10^7$MCS, we record the time trajectory $\{l_n = l(t=t_n)\}$ of the PI length at discrete time steps $t_n = n \tau_0$ with $n \in \mathbb{Z}$ and $\tau_0=10^3$. To reduce an artifact from short-time fluctuations, $l_n$ represents a running average over the time interval $t_{n-n_0} < t < t_n$ with $n_0=10$.
    
    \item  Given a trajectory $\{l_n\}$, we count $N_{l\to l\pm 1}$, the number of jumps in $l_n$ from $l$ to $l\pm 1$, and $T_l$, the total time span in which $l_n=l$.  The growth rate $W_g(l)$ and the shrink rate $W_s(l)$, per unit MCS, are given by $N_{l\to l+1}/ T_l$ and $N_{l\to l-1}/T_l$, respectively, which are then averaged over more than $10^3$ independent trajectories.
\end{enumerate} 

\begin{figure}
    \includegraphics[width=\columnwidth]{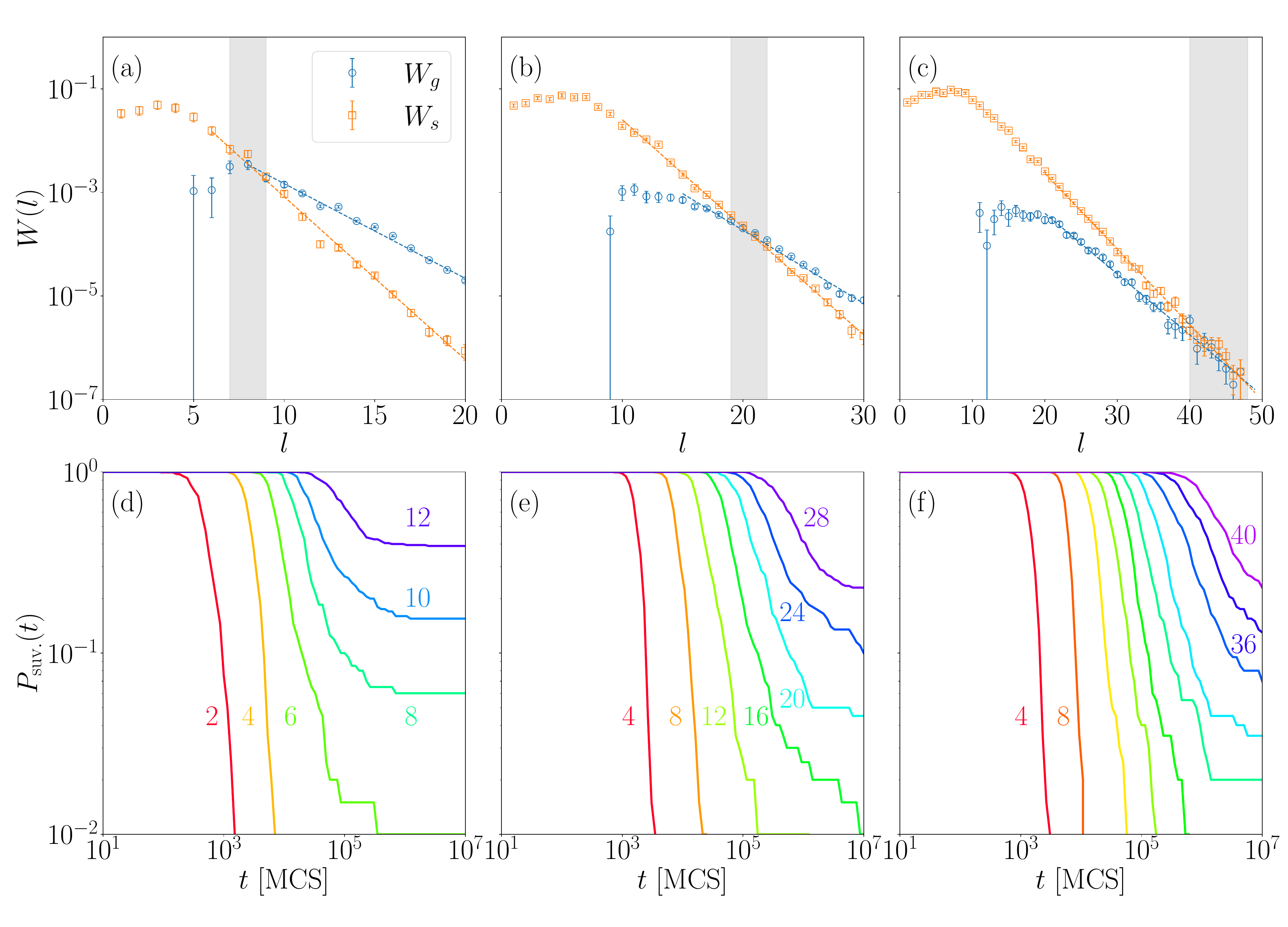}
    \caption{Growth ($W_g$) and shrink ($W_s$) rates per unit MCS of an isolated PI of
        size $l$ for $D=0.5$ in (a), $0.8$ in (b), and $1.0$ in (c). The shaded vertical bars
represent the threshold value $l_{\rm th.}$ beyond which the growth is
dominant. Survival probability $P_{\rm suv.}(t)$ of an isolated PI imposed
by a droplet of diameter $l$ for $D=0.5$ in (d), $D=0.8$ in (e), and $D=1.0$
in (f). [Parameters:  $L_x=1024, L_y=128, v=1, \beta=5, \rho_0 = 4, \rho_0'=200$]} \label{figS1}
\end{figure}

The growth and shrink rates are presented in Fig.~\ref{figS1} (a-c). The numerical
data clearly demonstrate the exponential scaling law and that $l_g > l_s$
for the parameter values considered. We
also notice the existence of a threshold size $l_{\rm th.}$: shorter PIs
with $l < l_{\rm th.}$ tend to evaporate while longer PIs with $l>l_{\rm
th.}$ tend to grow. It is analogous to the critical nucleation size in the
equilibrium nucleation theory for supersaturated gas systems. The threshold
value increases as $D$ increases: $l_{\rm th.} \simeq 8.0$ at $D=0.5$,
$\simeq 20.0$ at $D=0.8$, and $\simeq 45.0$ at $D=1.0$. It explains the
reason why the system starting from OIC takes longer time to relax to the
MPIS.

An isolated PI with $l \ll l_{\rm th.}$ shrinks and evaporates eventually. 
We confirmed this phenomenon by measuring the survival probability 
$P_{\rm suv.}(t)$ of an isolated PI generated from a circular droplet of diameter $l$ 
embedded in a homogeneous ordered state of density $\rho_0$~(see Fig.~3 of the main text). The initial particle density of the droplet is taken to be $5\rho_0$. 
The survival probability is given by the fraction of samples with a surviving PI among
all independent $200$ samples. Figure~\ref{figS1} (d-f)
presents the plots of $P_{\rm suv.}(t)$ in the log-log scale. 
For small $l$, the long time tail of $P_{\rm suv.}(t)$ is
characterized by a downward curvature. However, the tail
becomes flatter as $l$ increases, and attains an upward curvature
for large $l$. The crossover occurs around the threshold size $l_{\rm
th.}$. Thus, we expect that the PI larger than the threshold size has a finite
survival probability. 

\end{document}